# On the variational formulation of systems with non-holonomic constraints

**D. H. Delphenich**[†]

Lindsborg, KS 67456 USA

In a previous article by the author, it was shown that one could effectively give a variational formulation to non-conservative mechanical systems by starting with the first variation functional instead of an action functional. In this article, it is shown that this same approach will also allow one to give a variational formulation to systems with non-holonomic constraints. The key is to use an adapted anholonomic local frame field in the formulation, which then implies the replacement of ordinary derivatives with covariant ones. The method is then applied to the case of a vertical disc rolling without slipping or friction on a plane.

## 1    Introduction

Although Hamilton's principle of least action is far-reaching in its profundity in the eyes of physics, as well as other disciplines, it is not without it limitations. In effect, the action functional that one starts with is a sort of "infinite-dimensional" potential function, which means that when one is dealing with mechanical systems in which non-potential sorts of effects are present, there is a fundamental error being made in the formulation by trying to work around the non-existence of potential functions.

For instance, when the forces acting on a system include dissipative forces, such as friction, not all of the forces that contribute to the resultant will be represented by exact 1-forms on the configuration manifold. Another example is given by non-holonomic forces of constraint, while a third example is given by open systems in which one must account for "external" influences on the system.

Previously [**1**], the author addressed the variational formulation of non-conservative systems and showed that since the extremals of the action functional $S[x]$ are more fundamentally related to the zeroes of the *first variation functional* $\delta S[\delta x]$, it is more appropriate to regard that latter functional as the fundamental one. Its integrand is defined by a fundamental 1-form:

$$\phi = F_\mu dx^\mu + \Pi^a_\mu dx^\mu_a \tag{1.1}$$

---

[†] E-mail: david_delphenich@yahoo.com

on the manifold $J^1(\mathcal{O}; M)$ of 1-jets of local differentiable functions $x: \mathcal{O} \to M$; here, $\mathcal{O}$ is an $r$-dimensional parameter manifold with boundary and $M$ is an $m$-dimensional configuration manifold. The first term in f represents the generalizes forces that act on the system, while the second term represents the generalized momenta, which can include infinitesimal stresses. One then regards the action functional as a special case scenario in which the fundamental 1-form $\phi$ is exact and representable as $d\mathcal{L}$ for some Lagrangian density function $\mathcal{L}$ on $J^1(\mathcal{O}; M)$.

The key to obtaining equations of motion from the first variation functional is to recognize that it depends upon the vector field $\delta x$ on the image of $x$ by the intermediary of its 1-jet prolongation $\delta^1 x$; such a vector field on $J^1(\mathcal{O}; M)$ is then *integrable*. When one evaluates $\phi(\delta^1 x)$ one then finds that precisely the same steps that one uses when $\phi = d\mathcal{L}$ lead to an integrand of the form:

$$D^* \phi(\delta x) \mathcal{V} = \left[ \left( F_\mu - \frac{\partial \Pi_\mu^a}{\partial t^a} \right) \delta x^\mu \right] \mathcal{V} \tag{1.2}$$

that is integrated over $\mathcal{O}$ and another one that is integrated over $\partial \mathcal{O}$; With the usual boundary conditions on $\delta x$, this gives equations of motion in the form:

$$D^* \phi = \left( F_\mu - \frac{\partial \Pi_\mu^a}{\partial t^a} \right) dx^\mu = 0, \tag{1.3}$$

and when $\phi = d\mathcal{L}$, one finds that one has, indeed:

$$D^* \phi = \frac{\delta \mathcal{L}}{\delta x}, \tag{1.4}$$

since:

$$F_\mu = \frac{\partial \mathcal{L}}{\partial x^\mu}, \qquad \Pi_a^\mu = \frac{\partial \mathcal{L}}{\partial x_a^\mu} \tag{1.5}$$

when $\phi = d\mathcal{L}$.

One sees that, in effect, one is really using a generalization of d'Alembert's principle of virtual work (cf., e.g., Lanczos [2]) in place of Hamilton's principle of least action as the basis for the dynamical principle, since $D^* \phi(\delta x)$ takes the form of the virtual work that is done when the variation $\delta x$ is regarded as a virtual displacement of the object in $M$ that is described by the map $x$.

This fact is particularly relevant to the present investigation since the power of d'Alembert's principle is felt most acutely in the context of constrained motions. In the variational formulation, one finds that what one is doing is to restrict the set of physically acceptable variations $\delta x$ to not just the ones that vanish on the boundary of $x$ or satisfy the transversality condition for a free-boundary variational problem, but also must lie in some specified submanifold $C_x M$ of each tangent space $T_x M$ to each $x \in M$. In the following discussion, we shall confine ourselves to the most commonly encountered case of submanifolds that take the form of linear subspaces, but the method can be generalized.

The set $C(M)$ of all these constraint subspaces then defines a vector sub-bundle of the vector bundle $T(M)$, and the issue of holonomy versus non-holonomy is that of the integrability of that sub-bundle, respectively. When $C(M)$ is integrable there is a foliation of $M$ by leaves whose dimension equals the dimension of each $C_x$, which we will assume to be the same everywhere on $M$. The imposition of holonomic constraints then reduces the basic variational problem to a corresponding problem on each leaf. In the non-integrable case, no such foliation exists, and one sees that the dimension of each constraint space $C_x M$ will be strictly less than the dimension of the configuration manifold, which is still all of $M$, in general.

As we will see, when one starts with the first variation functional all that is necessary in order to account for non-holonomic constraints in one's basic variational problem of finding the zeroes of the first variation functional is to use anholonomic [1] local frame fields that are adapted to the sub-bundle $C(M)$ and replace the ordinary partial derivatives $\partial_\mu$ with respect to the coordinates $x^\mu$ of a chart on $M$ with "generalized covariant derivatives" $\nabla_\mu$ with respect to a generalized connection that is defined by the adapted local frame field.

In the next section, we shall discuss the geometrical nature of constrained configuration spaces, while in the following section we shall apply these considerations to the problem of obtaining variational equations of motion for constrained mechanical system. In section 4, we shall show how everything applies to a common problem of a mechanical system with a linear non-holonomic constraint, namely, a disc rolling with slipping or dissipative forces on a plane. Finally, we discuss how the same approach of starting with the first variation functional might be applied to systems with external influences, such as the measurements that act on quantum systems.

## 2        Linearly constrained mechanical systems

The mechanical systems that we shall be dealing with will have states that take the form of objects defined by differentiable maps $x \colon \mathcal{O} \to M$, $t \mapsto x(t)$ from an $r$-dimensional parameter manifold $\mathcal{O}$ that has a boundary $\partial \mathcal{O}$ into an $m$-dimensional configuration manifold $M$; we shall, moreover, assume that $r \leq m$. If $(U, t^a)$ is a coordinate chart on some open subset $U \subset \mathcal{O}$ and $(V, x^\mu)$ is a chart on some open subset $V \subset M$ that contains the image $x(\mathcal{O})$ then one can express this map in the form of a system of equations of the form:

---

[1] It may sound inconsistent to use the word "non-holonomic" to describe the constraint and "anholonomic" to describe the adapted local frame field when they are so clearly related, but the use of both terms in their respective fields of application – viz., constrained mechanics and the geometry of moving frames – is so well-established that we shall simply use both terms as the situation dictates.

$$x^\mu = x^\mu(t^a). \tag{2.1}$$

One can then regard the map $x$ as defining a submanifold of $M$.

At each point $t \in \mathcal{O}$, one can define the differential map $dx|_t : T_t\mathcal{O} \to T_{x(t)}M$, which is a linear map from the $r$-dimensional vector space $T_t\mathcal{O}$ into the $m$-dimensional vector space $T_{x(t)}M$. Its local expression in terms of the chosen coordinate systems is then the matrix $x^\mu{}_{,a} = \partial x^\mu / \partial t^a$. It will be convenient in what follows to also regard $dx|_t$ as an element of the tensor product space $T_t^*\mathcal{O} \otimes T_{x(t)}M$, and express it in the local form:

$$dx|_t = x^\mu{}_{,a}\, dt^a \otimes \partial_\mu\,. \tag{2.2}$$

One must be aware that unless one restricts the nature of the map $x$ suitably the differential map will not necessarily map the $r$-dimensional vector space $T_t\mathcal{O}$ to an $r$-dimensional subspace of $T_{x(t)}M$. This is true iff the rank of $dx|_t$ is $r$, which is equivalent to saying that it is a linear injection. If this is the case for all $t \in \mathcal{O}$ then, by definition, $x$ defines an *immersed* submanifold of $M$. Such submanifolds can still have singularities, such as self-intersections, so a stronger condition is to make $x$ an *embedded* submanifold of $M$, which also makes $x$ a homeomorphism onto its image $x(\mathcal{O})$, when one gives that subspace the topology whose open subsets are all intersections of open subsets of $M$ with itself $x(\mathcal{O})$. However, if one is dealing locally with points of $x(\mathcal{O})$ and their tangent objects, the distinction between an immersion and an embedding is not significant, since that distinction has a more global character.

Examples of mechanical systems that can be expressed in the form that we have chosen include differentiable curve segments, static surfaces or moving filaments, static volumes or moving membranes, moving rigid bodies, and harmonic oscillators. In the case of dynamical systems, one can include time as one of the parameters in $\mathcal{O}$. One of the advantages of the d'Alembertian formulation of dynamics – in addition to its significance for constrained mechanical systems – is the fact that it allows one to treat dynamical problems as essentially static problems in a higher-dimensional space.

Mechanical systems can generally have two basic types of constraints on their kinematical variables: constraints on the possible positions – or, more generally, *states* – of the system and constraints on the parameter derivatives $x^\mu{}_{,a}$ of these states. The first type of constraint is already accounted for the definition of the configuration manifold $M$, which might take the form of a region with boundary in $\mathbf{R}^m$ or perhaps the torus $T^m$, in the case where the position variables are independent angles.

Constraints on the parameter derivatives $x^\mu{}_{,a}$ generally take the form of specifying that all $m$ of the tangent vectors:

$$\mathbf{x}_a = x^\mu{}_{,a}\, \partial_\mu\,, a = 1, \ldots, r \tag{2.3}$$

in each $T_{x(t)}M$ for each $t \in \mathcal{O}$ lie in some specified subspace $C_{x(t)}M$ of $T_{x(t)}M$.

One immediately classifies the constraint as *linear* or *nonlinear* according to whether the subspace $C_xM$ is a linear subspace at every point of $x$ or not.

Most commonly, in the linear case one defines the constraint spaces to be the annihilating subspaces of some linearly independent set of local 1-forms $\theta^\alpha$, $\alpha = N+1, \ldots, m$ on some open subspace $U \subset M$, where $N$ is the dimension of any vector space $C_xM$; i.e. for any tangent vector $\mathbf{v} \in C_xM$, one must have:

$$\theta^\alpha(\mathbf{v}) = 0, \qquad \alpha = N+1, \ldots, m. \tag{2.4}$$

Typically, one would complete the $m$-$N$-coframe $\theta^\alpha$ to an $m$-frame $\theta^\mu$, $\mu = 1, \ldots, m$ by means of $N$ complementary linearly independent 1-forms $\theta^i$, $i = 1, \ldots, N$ on $U$. Its reciprocal local frame field on $U$ consists of the $m$ linearly independent local vector fields $\mathbf{e}_\mu$, $\mu = 1, \ldots, m$ that are defined by:

$$\theta^\mu(\mathbf{e}_\nu) = \delta^\mu{}_\nu. \tag{2.5}$$

One refers to such a local frame field and its reciprocal coframe field as being *adapted* to the linear subspaces that are defined by the fibers of $C(M)$.

Since the components of $\mathbf{v}$ with respect to the frame $\mathbf{e}_\mu$ are:

$$v^\mu = \theta^\mu(\mathbf{v}) \tag{2.6}$$

one can also express the constraint in the forms:

$$v^\alpha = 0, \qquad \alpha = N+1, \ldots, m, \tag{2.7}$$

or:

$$\mathbf{v} = v^i \mathbf{e}_i, \qquad i = 1, \ldots, N. \tag{2.8}$$

A typical nonlinear constraint might take the form of requiring that the $\mathbf{b}_a$ always lie on the surface of a sphere in each tangent space. However, we shall consider only the linear case in the sequel.

One can also distinguish constraint sub-bundles according to whether they are integrable or not. One then refers to the constraints as *holonomic* or *non-holonomic*, respectively, which we clarify in the following subsections, after we briefly summarize the geometry of adapted local frame fields. In particular, we discuss a useful example of a non-holonomic constraint that is called *semi-holonomic*, and is based in the way that Chaplygin represented linear non-holonomic constraints (cf., e.g., Neimark and Fufaev [**3**] on this).

### 2.1    Adapted local frame fields

Suppose $C(M)$ is a vector sub-bundle of $T(M)$ that has constant rank $N$; i.e., its fibers have the same dimension $N$ everywhere. A local frame field $\{\mathbf{e}_\mu, \mu = 1, \ldots, m\}$ on an open subset $U \subset M$ is said to be *adapted* to the sub-bundle $C(M)$ iff one can partition it into to subsets $\{\mathbf{e}_i, i = 1, \ldots, N\}$ and $\{\mathbf{e}_\alpha, \alpha = N+1, \ldots, m\}$ such that the former $N$-frame spans $C_x U$ at each point $x \in U$. Although the latter subset spans a complementary $m$-$N$-dimensional subspace, as well, that fact is usually regard as irrelevant, unless one has been given a complementary sub-bundle to $C(M)$.

The reciprocal coframe field $\theta^\mu$ to $\mathbf{e}_\mu$ will then have the property that the subspaces $C_x U$ are annihilated by $m-N$-coframe field $\{\theta^\alpha, \ \alpha = N+1, \ \ldots, \ m\}$ on $U$, while the complementary subspaces that are spanned by $\{\mathbf{e}_\alpha, \ \alpha = N+1, \ \ldots, \ m\}$ will be annihilated by the $\{\theta^i, i = 1, \ \ldots, N\}$.

A general notion that we need to introduce at this point is the concept of the *structure functions* of a local frame field. Since $\mathbf{e}_\mu$ define a basis for each tangent space to $U$, any tangent vector – including $[\mathbf{e}_\mu, \mathbf{e}_\nu]$ for each combination of $\mu$ and $\nu$ – can be expressed as a linear combination of the frame vectors:

$$[\mathbf{e}_\mu, \mathbf{e}_\nu] = c_{\mu\nu}^\lambda \, \mathbf{e}_\lambda \, . \tag{2.9}$$

Since the $\mathbf{e}_\mu$ are functions of the points of $U$ the coefficients $c_{\mu\nu}^\lambda$ will be functions of those points, as well, and one calls them the *structure functions* of the local frame field $\mathbf{e}_\mu$. In the case where $M$ is the manifold of a Lie group $G$ and the frame field is left-invariant, they will be constants, and indeed they will be the structure constants of the Lie algebra of $G$, as they are usually defined (see Sattinger and Weaver [**4**], for example, on this).

The natural local frame field $\{\partial_\mu, \ \mu = 1, \ \ldots, \ m\}$ that is associated with any coordinate chart $(U, x^\mu)$ has the property that the Lie subalgebra of $\mathfrak{X}(U)$ that it spans is Abelian. Hence:

$$[\mathbf{e}_\mu, \mathbf{e}_\nu] = 0; \tag{2.10}$$

i.e.:

$$c_{\mu\nu}^\lambda = 0. \tag{2.11}$$

Such a local frame field is called *holonomic*; otherwise, one calls it *anholonomic*.

When $U$ is given a coordinate system $x^\mu$ one can always express an arbitrary local frame field $\mathbf{e}_\mu$ by means of a transition function $A: U \rightarrow GL(m)$:

$$\mathbf{e}_\mu(x) = A_\mu^\nu(x)\partial_\nu \, , \tag{2.12}$$

In such a representation, the structure functions can be obtained in terms of the partial derivatives of $A$ by direct evaluation of (2.9) using (2.12):

$$c_{\mu\nu}^\lambda = (A_\mu^\kappa \partial_\kappa A_\nu^\rho - A_\nu^\kappa \partial_\kappa A_\mu^\rho)\tilde{A}_\rho^\lambda \, , \tag{2.13}$$

in which the tilde over the matrix signifies that one is using the inverse of that matrix.

When the local frame field $\mathbf{e}_\mu$ is adapted to $C(U)$ one can partition its elements into $\{\mathbf{e}_i, \ i = 1, \ \ldots, \ N\}$ and $\{\mathbf{e}_\alpha, \ \alpha = N+1, \ \ldots, \ m\}$, and then further partition the commutation relations for the local frame field into:

$$[\mathbf{e}_i\,,\,\mathbf{e}_j] = c_{ij}^k \mathbf{e}_k + c_{ij}^\alpha \mathbf{e}_\alpha\,, \tag{2.14a}$$

$$[\mathbf{e}_i\,,\,\mathbf{e}_\alpha] = c_{i\alpha}^k \mathbf{e}_k + c_{i\alpha}^\beta \mathbf{e}_\beta\,, \tag{2.14b}$$

$$[\mathbf{e}_\alpha\,,\,\mathbf{e}_\beta] = c_{\alpha\beta}^k \mathbf{e}_k + c_{\alpha\beta}^\gamma \mathbf{e}_\gamma\,. \tag{2.14c}$$

One can formulate the commutation relations for $\mathbf{e}_\mu$ in terms of its reciprocal coframe field $\theta^\mu$ by using the "intrinsic" formula for the exterior derivative[1] $d\theta^\mu$ :

$$d\theta^\lambda(\mathbf{e}_\mu\,,\,\mathbf{e}_\nu) = \mathbf{e}_\mu\,\theta^\lambda(\mathbf{e}_\nu) - \mathbf{e}_\nu\,\theta^\lambda\,(\mathbf{e}_\mu) - \theta^\lambda\,([\mathbf{e}_\mu\,,\,\mathbf{e}_\nu]). \tag{2.15}$$

The first two terms vanish since the coframe field is reciprocal, and by introducing the structure functions in the last term, one finds:

$$d\theta^\lambda(\mathbf{e}_\mu\,,\,\mathbf{e}_\nu) = -\,c_{\mu\nu}^\lambda\,; \tag{2.16}$$

hence:

$$d\theta^\lambda = -\tfrac{1}{2} c_{\mu\nu}^\lambda \theta^\mu \wedge \theta^\nu\,. \tag{2.17}$$

In the case where $M$ is a Lie group manifold and $\mathbf{e}_\mu$ is left-invariant these become the Maurer-Cartan equations.

When $\mathbf{e}_\mu$ is holonomic one also must have:

$$d\theta^\mu = 0, \qquad \text{all } \mu = 1, \dots, m. \tag{2.18}$$

If $\theta^\mu$ is adapted to $C(U)$ – so each $C_x U$ is annihilated by all $\theta^\alpha$, $\alpha = N+1, \dots, m$ – then one can partition it into $\{\theta^i, i = 1, \dots, N\}$ and $\{\theta^\alpha, \alpha = N+1, \dots, m\,\}$, which then allows one to also partition (2.17) into:

$$d\theta^i = -\tfrac{1}{2} c_{jk}^i \theta^j \wedge \theta^k + \tfrac{1}{2} c_{j\beta}^i \theta^j \wedge \theta^\beta + \tfrac{1}{2} c_{\beta\gamma}^i \theta^\beta \wedge \theta^\gamma\,. \tag{2.19a}$$

$$d\theta^\alpha = -\tfrac{1}{2} c_{jk}^\alpha \theta^j \wedge \theta^k + \tfrac{1}{2} c_{j\beta}^\alpha \theta^j \wedge \theta^\beta + \tfrac{1}{2} c_{\beta\gamma}^\alpha \theta^\beta \wedge \theta^\gamma\,. \tag{2.19b}$$

When $\mathbf{e}_\mu$ is related to $\partial_\mu$ as in (2.12) one then has that $\theta^\mu$ is related to $dx^\mu$ by way of:

$$\theta^\mu = \tilde{A}_\nu^\mu dx^\nu \tag{2.20}$$

since that coframe field must be reciprocal to the frame field $\mathbf{e}_\mu$.

## 2.2 Holonomic constraints

When one is given a vector sub-bundle of $T(M)$, such as $C(M)$, with $N$-dimensional fibers one can always regard such a sub-bundle as a *differential system* on M  The question then arises whether there is some unique embedded $N$-dimensional submanifold

---

[1] This notion is discussed in any book that covers the basics of exterior differential forms, such as Warner [**5**].

$\iota: L \to M$ that passes through each point $x \in M$ and has a tangent space[1] $T_x(L)$ that coincides with $C_x(M)$. Such a submanifold is called an *integral submanifold* of the differential system $C(M)$ and the set of all such integral submanifolds defines a partition of $M$ into disjoint integral submanifolds. Such a partition is called a *foliation* of $M$ of dimension $N$ and the individual integral submanifolds are called its *leaves*.

The necessary and sufficient condition for the complete integrability[2] of $C(M)$ is that if $\mathbf{X}, \mathbf{Y} \in \mathfrak{X}(C)$, which is the vector space of all vector fields on $M$ that take their values in the fibers of $C(M)$, then $[\mathbf{X}, \mathbf{Y}]$ is also a vector field in $\mathfrak{X}(C)$; i.e., $\mathfrak{X}(C)$ is a Lie subalgebra of the Lie algebra $\mathfrak{X}(M)$ of all vector fields on $M$.

If $\mathbf{e}_\mu$, $\mu = 1, \ldots, m$ is a local frame field on $U \subset M$ that is adapted to $C(U)$ then one can express $\mathbf{X}$ as $X^i \mathbf{e}_i$ and $\mathbf{Y}$ as $Y^j \mathbf{e}_j$, which makes:

$$[\mathbf{X}, \mathbf{Y}] = X^i Y^j [\mathbf{e}_i, \mathbf{e}_j]. \tag{2.21}$$

Hence, the real issue is whether $[\mathbf{e}_i, \mathbf{e}_j]$ is a vector in the subspace spanned by the $\mathbf{e}_i$ for each possible value of $i$ and $j$. Using the structure functions for $\mathbf{e}_\mu$, one sees that the differential system $C(U)$ is completely integrable on $U$ iff:

$$c_{ij}^\alpha = 0 \qquad \text{all } i, j = 1, \ldots, N, \ \alpha = N+1, \ldots, m. \tag{2.22}$$

In such an event, one can find a – generally non-constant – frame transition function $A: U \to GL(N)$, $x \mapsto A_j^i(x)$ that makes:

$$\mathbf{e}_i(x) = A_j^i(x) \partial_i \tag{2.23}$$

for some natural $N$-frame field $\partial_i$ that is defined by a coordinate chart $(U, x^\mu)$ on $U$ that is adapted to $C(U)$.

Hence, although complete integrability of the constraint sub-bundle $C(U)$ does not demand that the chosen adapted local frame field be holonomic, nonetheless, when $C(M)$ is completely integrable there will always be a change of adapted frame field that takes the $N$-frame $\mathbf{e}_i$ that is tangent to the leaves of the foliation of $U$ to a natural – hence, holonomic – local frame field.

The subtlety that is hiding in this statement is the fact that although when $U$ is given a coordinate system $x^\mu$ one can *always* express $\mathbf{e}_\mu$ by means of a transition function $A: U \to GL(m)$, as in (2.12), nevertheless, it is only when $U$ is foliated by integral submanifolds in the manner above that one can restrict the values of $A$ to a $GL(N)$ subgroup in such a manner that the subspaces $C_x U$ are preserved at each $x \in U$ under the transition.

When the constraint $C(M)$ is holonomic, one can say that the number of *degrees of freedom* in the mechanical system – i.e., the dimension $N$ of each fiber of $C(M)$ – is equal to the dimension of its configuration space of allowable positions – namely, the leaf through each point. Hence, the concept of degrees of freedom has more to do with the allowable directions of velocity vectors than it does with allowable positions.

---

[1] By this, we mean the image of the corresponding tangent space to $L$ under the map $\iota$ and not the tangent to $M$ at the point $x$; the former will be of dimension $N$, while the latter has dimension $m$.

[2] The use of the word "complete" implies that the integral submanifolds have maximal dimension $N$. It is possible that integral submanifolds of lower dimension might exist, more generally, though.

### 2.3    Non-holonomic constraints

If the constraint sub-bundle $C(M)$ is non-holonomic then this latter situation is no longer true. In general, the number of degrees of freedom will be strictly less than the dimension of the configuration space.

Any adapted local frame field to such a $C(M)$ will have to satisfy:

$$c_{ij}^\alpha \neq 0 \qquad \text{for some } i, j = 1, \ldots, N, \ \alpha = N+1, \ldots, m. \tag{2.24}$$

Since the complementary linear subspace to $C_x U$ that is spanned by the $\mathbf{e}_\alpha$, $\alpha = N+1$, ..., $m$ is somewhat irrelevant, a useful choice of anholonomic local frame field for the purposes of non-holonomic mechanical constraints is given the *semi-holonomic* local frame fields. For such local frame fields, the $\mathbf{e}_\alpha$ all commute amongst themselves:

$$[\mathbf{e}_\alpha, \mathbf{e}_\beta] = 0 \qquad \text{for all } \alpha, \beta = N+1, \ldots, m. \tag{2.25}$$

This makes:

$$c_{\alpha\beta}^i = c_{\alpha\beta}^\gamma = 0 \qquad \text{for all } i = 1, \ldots, N, \ \alpha, \beta, \gamma = N+1, \ldots, m. \tag{2.26}$$

Hence, it is no loss of generality to set:

$$\mathbf{e}_\alpha = \partial_\alpha \tag{2.27}$$

for some coordinate system $(U, x^\mu)$.

If one considers linear constraints of the Chaplygin form:

$$v^i \text{ arbitrary}, \qquad v^\alpha = a_i^\alpha v^i, \tag{2.28}$$

relative to a natural local frame field, then this suggests that one should define:

$$\theta^i = dx^i, \qquad \theta^\alpha = -a_i^\alpha dx^i + dx^\alpha, \tag{2.29}$$

in order to make the constraint take the form:

$$\theta^\alpha(\mathbf{v}) = 0. \tag{2.30}$$

From (2.29), one can then express the matrix $\tilde{A}_\nu^\mu$ that appears in (2.20) in block form:

$$\tilde{A}_\nu^\mu = \left[ \begin{array}{c|c} \delta_j^i & 0 \\ \hline -a_j^\alpha & \delta_\beta^\alpha \end{array} \right].$$
(2.31)

The inverse of this matrix, which is obtained by simply solving (2.29) for the $dx^i$ and $dx^\alpha$, is:

$$A_\nu^\mu = \left[ \begin{array}{c|c} \delta_j^i & 0 \\ \hline a_j^\alpha & \delta_\beta^\alpha \end{array} \right],$$
(2.32)

which is of the same block-matrix form.

Furthermore, if one multiplies $A_\nu^\mu$ by another matrix $B_\nu^\mu$ that has the same block-matrix form, except that the non-trivial sub-matrix is $b_j^\alpha$, then the result is:

$$[AB]_\nu^\mu = \left[ \begin{array}{c|c} \delta_j^i & 0 \\ \hline a_j^\alpha + b_j^\alpha & \delta_\beta^\alpha \end{array} \right].$$
(2.33)

Therefore, the set of all matrices of the form (2.32) represents a Lie group that is isomorphic to the group of translations of $\mathbf{R}^{N(m-N)}$.

Using (2.32), we can express the local semi-holonomic frame field that is reciprocal to $\theta^\mu$ in the form:

$$\mathbf{e}_i = \partial_i + a_i^\alpha \partial_\alpha, \qquad \mathbf{e}_\alpha = \partial_\alpha.$$
(2.34)

In order to obtain the structure functions for this local frame field, we can use either (2.29) or (2.34); we choose the former and obtain:

$$d\theta^i = 0,$$
(2.35a)
$$d\theta^\alpha = -\tfrac{1}{2}(a_{i,j}^\alpha + a_{i,\beta}^\alpha a_j^\beta - a_{j,i}^\alpha - a_{j,\beta}^\alpha a_i^\beta)\theta^i \wedge \theta^j.$$
(2.35b)

The form of the expression in parentheses is particularly interesting, since it takes the form of an anti-symmetrized total derivative of the matrix $a_i^\alpha$ with respect to $x^i$, under the assumptions that $x^\alpha$ is a function of $x^i$ and $a_i^\alpha = \partial x^\alpha / \partial x^i$.

From (2.35a, b), we obtain the non-zero structure functions by inspection:

$$c_{ij}^\alpha = a_{i,j}^\alpha + a_{i,\beta}^\alpha a_j^\beta - a_{j,i}^\alpha - a_{j,\beta}^\alpha a_i^\beta.$$
(2.36)

## 2.4   Derivatives in anholonomic frames

Chaplygin, and later Neimark and Fufaev [**3**], attempted to give a variational formulation to mechanical systems with non-holonomic constraints by the introduction of what they were calling "quasi-coordinates." One can see that what these quasi-coordinates actually represented were the path-dependent functions on configuration space that one obtains by integrating 1-forms that are not exact along paths that lead away from some chosen initial point. Hence, in the present context, it becomes clear that it is preferable to deal with the non-exact 1-forms directly, which is what we are doing by considering the 1-forms $\theta^\alpha$ to be the fundamental objects.

The point at which one needs to be especially careful is in the treatment of differentials and derivatives when one is looking at their components with respect to anholonomic frame fields.

These considerations are not relevant when one is considering the differential $df$ of a differential function $f$ on $M$, since the local components of $df$ transform from a holonomic local coframe field $dx^\mu$ to an anholonomic one $\theta^\mu = A^\mu_\nu dx^\nu$ by way of the inverse of the transition function $A$:

$$df = f_{,\mu}\, dx^\mu = f_{,\nu}\, \tilde{A}^\nu_\mu \theta^\mu = \overline{f}_{,\mu} \theta^\mu, \tag{2.37}$$

which means that the transformation of components is simply:

$$\overline{f}_{,\mu} = \tilde{A}^\nu_\mu f_{,\nu}. \tag{2.38}$$

However, when one goes to vector fields, covector fields, and higher-rank tensor fields, one must inevitably differentiate the transition function itself and this adds a contribution to the transformed components. For instance, for a vector field:

$$\mathbf{X} = X^\mu \partial_\mu = \overline{X}^\mu \mathbf{e}_\mu = (X^\nu \tilde{A}^\mu_\nu)\mathbf{e}_\mu \tag{2.39}$$

one finds that:

$$d\mathbf{X} = dX^\mu \otimes \partial_\mu = X^\mu{}_{,\nu}\, dx^\nu \otimes \partial_\mu \tag{2.40}$$

relative to the holonomic frame field, while:

$$d\mathbf{X} = d\overline{X}^\mu \otimes \mathbf{e}_\mu + \overline{X}^\mu \otimes d\mathbf{e}_\mu \tag{2.41}$$

relative to the anholonomic one, since the differential of the local frame field $\mathbf{e}_\mu$ does not, presumably, vanish identically.

In order to evaluate this differential, one can express $\mathbf{e}_\mu$ in terms of $\partial_\mu$ by means of the transition function $A$ and differentiate the components:

$$d\mathbf{e}_\mu = dA^\nu_\mu \otimes \partial_\nu = dA^\lambda_\mu \tilde{A}^\nu_\lambda \otimes \mathbf{e}_\nu \equiv -\gamma^\nu_\mu \otimes \mathbf{e}_\nu, \tag{2.42}$$

in which we have introduced the local 1-form:

$$\gamma^{\nu}_{\mu} = -\,dA^{\lambda}_{\mu}\tilde{A}^{\nu}_{\lambda} = -A^{\lambda}_{\mu,\kappa}\tilde{A}^{\nu}_{\lambda}dx^{\kappa} = -(A^{\lambda}_{\mu,\rho}\tilde{A}^{\nu}_{\lambda}\tilde{A}^{\rho}_{\kappa})\theta^{\kappa} \equiv \gamma^{\nu}_{\mu\kappa}\theta^{\kappa}\,, \tag{2.43}$$

which can be regarded as either a matrix of local 1-forms or a local form that takes its values in the Lie algebra $\mathfrak{gl}(m)$ to the Lie group $GL(m)$. In particular, we single out the definition in (2.43):

$$\gamma^{\nu}_{\mu\kappa} = A^{\lambda}_{\mu,\rho}\tilde{A}^{\nu}_{\lambda}\tilde{A}^{\rho}_{\kappa}\,, \tag{2.44}$$

which then exhibits the fact that the anti-symmetrized components of $\gamma^{\nu}_{\mu}$ are the structure functions of the local frame field that we are using.

We can now put (2.41) into the form:

$$d\mathbf{X} = (d\overline{X}^{\mu} + \gamma^{\mu}_{\nu}\overline{X}^{\nu}) \otimes \mathbf{e}_{\mu} \equiv \nabla\overline{X}^{\mu} \otimes \mathbf{e}_{\mu}\,, \tag{2.45}$$

in which we have introduced the local *covariant differential* of the components $\overline{X}^{\mu}$:

$$\nabla\overline{X}^{\mu} = d\overline{X}^{\mu} + \gamma^{\mu}_{\nu}\overline{X}^{\nu}\,. \tag{2.46}$$

Since any discussion of differentials of local frame fields must inevitably come around to the subject of local connections, we point out that what we have defined above in the form of $\gamma^{\nu}_{\mu}$ is a local linear connection that is canonically associated with the local frame field in question. One can think of it as the "teleparallelism" connection that makes the local frame field parallel. In general, this canonical local linear connection cannot be assembled into a global linear connection on the bundle $GL(M)$ of linear frames on $M$ in a canonical way, unless $M$ is itself parallelizable; i.e., unless $M$ admits a global frame field. One can, however, assemble such local connections into a global linear connection in a non-canonical way by means of a partition of unity (see, Sternberg [**6**] on this).

When one introduces a more general linear connection $\omega^{\nu}_{\mu}$ on the local frame field $\mathbf{e}_{\mu}$, one cannot regard $\omega^{\nu}_{\mu}\mathbf{e}_{\nu}$ as the *definition* of $d\mathbf{e}_{\mu}$, since that geometrical object is defined independently of the introduction of any connection, and, as we have seen, implies only the introduction of the connection $\gamma^{\nu}_{\mu}$. Rather, one must *substitute* $\omega^{\nu}_{\mu}\mathbf{e}_{\nu}$ for $d\mathbf{e}_{\mu}$ in the differentiation of any local tensor field.

In order to find the components of the differential of a local covector field:

$$\alpha = \alpha_{\mu}\,dx^{\mu} = \overline{\alpha}_{\mu}\theta^{\mu} = (\alpha_{\nu}A^{\nu}_{\mu})\theta^{\mu} \tag{2.47}$$

we see that the only essential differences between (2.47) and (2.39) are in the replacement of $\tilde{A}^{\mu}_{\nu}$ with $A^{\mu}_{\nu}$ and $\mathbf{e}_{\mu}$ with its reciprocal $\theta^{\mu}$. Under differentiation, this has the effect of replacing $\gamma^{\nu}_{\mu}$ with $-\gamma^{\nu}_{\mu}$, since the differentiation of $\tilde{A}^{\mu}_{\kappa}A^{\kappa}_{\nu} = \delta^{\mu}_{\nu}$ gives:

$$d\tilde{A}_\kappa^{\ \mu} A_\nu^{\ \kappa} + \tilde{A}_\kappa^{\ \mu} dA_\nu^{\ \kappa} = 0. \tag{2.48}$$

Hence, one finds that:

$$d\alpha = d\alpha_\mu \otimes dx^\mu = \nabla \bar{\alpha}_\mu \otimes \theta^\mu, \tag{2.49}$$

in which we now have:

$$\nabla \bar{\alpha}_\mu = d\bar{\alpha}_\mu - \gamma_\mu^{\ \nu} \bar{\alpha}_\nu \tag{2.50}$$

as the dual analogue of (2.46).

The extension of the above to higher-rank tensor fields is immediate and based on the fact that the covariant derivative operator is a linear derivation with respect to the tensor product:

$$\nabla(T_1 \otimes T_2) = \nabla T_1 \otimes T_2 + T_1 \otimes \nabla T_2, \tag{2.51}$$

which extends to higher-rank tensor products by the associativity of the tensor product.

Since every local tensor field is a finite linear combination of tensor products of local vector fields and local covector fields with coefficients in the ring of smooth local functions, one can obtain the covariant derivative of any tensor field from the rules given above for functions, vector fields, and covector fields, by using (2.51) repeatedly.

One can also anti-symmetrize the covariant derivative of $k$-forms – i.e., completely anti-symmetric covariant tensor fields of rank $k$ – to obtain and *exterior covariant derivative:*

$$\nabla_\wedge \phi = d_\wedge \phi + \gamma^\wedge \phi. \tag{2.52}$$

This notation is really a shorthand for the more elaborate expression that pertains to the components of $d_\wedge \phi$. For instance, when $\phi = \phi_\mu \theta^\mu$ is a local 1-form, the components of $\nabla_\wedge \phi$ with respect to $\theta^\mu$ are most easily obtained from exterior differentiation:

$$d_\wedge \phi = d_\wedge \phi_\mu \wedge \theta^\mu + \phi_\mu \ d_\wedge \theta^\mu = \tfrac{1}{2}(\nabla_\wedge \phi)_{\mu\nu} \theta^\mu \wedge \theta^\nu. \tag{2.53}$$

When one uses the fact that:

$$d_\wedge \theta^\mu = \gamma_\nu^{\ \mu} \wedge \theta^\nu = -\tfrac{1}{2}(\gamma_{\kappa\nu}^{\ \ \mu} - \gamma_{\nu\kappa}^{\ \ \mu})\theta^\kappa \wedge \theta^\nu, \tag{2.54}$$

along with:

$$d_\wedge \theta^\mu = -\tfrac{1}{2} c_{\kappa\nu}^{\ \ \mu} \theta^\kappa \wedge \theta^\nu, \tag{2.55}$$

which we derived above, we see that we indeed have:

$$c_{\kappa\nu}^{\mu} = \gamma_{\kappa\nu}^{\mu} - \gamma_{\nu\kappa}^{\mu}. \tag{2.56}$$

This makes the components of the 2-form $d_\wedge\phi$ with respect to the local coframe field $\theta^\mu$ become:

$$(\nabla_\wedge\phi)_{\mu\nu} = -(\phi_{\mu,\,\nu} - \phi_{\nu,\,\mu} + c_{\mu\nu}^{\kappa}\,\phi_{\kappa}). \tag{2.57}$$

Since the set of 2-forms:

$$\Theta^\mu = \nabla_\wedge\theta^\mu = d_\wedge\theta^\mu + \gamma_\nu^\mu \wedge \theta^\nu = 2\,d_\wedge\theta^\mu \tag{2.58}$$

represents the torsion of the local linear connection $\gamma_\nu^\mu$, one sees that this torsion is essentially defined by the structure functions of the local coframe field that we are using. Had we used a connection 1-form with vanishing torsion, such as the Levi-Civita connection for some metric on $T(M)$, we would have found that in a holonomic coframe field the exterior covariant derivative of any $k$-form agrees with the exterior derivative.

For the sake of completeness, we also point out that the curvature 2-form of the connection 1-form $\gamma_\nu^\mu$ always vanishes:

$$\Omega_\nu^\mu \equiv \nabla_\wedge\gamma_\nu^\mu = d_\wedge\gamma_\nu^\mu + \gamma_\kappa^\mu \wedge \gamma_\nu^\kappa = 0. \tag{2.59}$$

This can be verified by direct calculation or by noting that the existence of a parallel local frame field is possible only if the curvature of the connection vanishes.

When $M$ is a Lie group manifold, and $\theta^\mu$ is left-invariant the connection form $\gamma_\nu^\mu$ that we defined is the Maurer-Cartan form, and the Maurer-Cartan equations can be obtained from (2.59), since the structure functions are constant. One can think of $\gamma_\nu^\mu$ as a generalized "intrinsic angular velocity" of the local frame field $\mathbf{e}_\mu$ or its reciprocal coframe field $\theta^\mu$, as this is how obtains the Coriolis corrections to the velocity and the Coriolis and centripetal corrections to the accelerations when they are measured in a rotating – i.e., anholonomic – local frame field in rotational mechanics.

When $\theta^\mu$ is adapted to a constraint sub-bundle $C(M)$ in the semi-holonomic manner that we have been describing all along, the only difference in the calculation of differential components amounts to the use of ordinary differentials and derivatives for the holonomic frame members and covariant differentials and derivatives for the anholonomic ones. For instance, if:

$$\beta = \beta_i\,dx^i + \beta_\alpha\theta^\alpha, \tag{2.60}$$

then:

$$d\beta = d\beta_i \otimes dx^i + \nabla\beta_\alpha \otimes \theta^\alpha. \tag{2.61}$$

### 3    Linearly constrained variations

Now that we have done the geometrical overhead of calculating the components of differentials of various geometric objects in anholonomic local frame fields, the inclusion of non-holonomic constraints into the formulation of variational problems in mechanics becomes quite straightforward. The two essential modifications are:

1. Defining the basic variational problem in terms of the first variation functional, instead of the usual action functional..

2. Expressing the mechanical variables in terms of an adapted anholonomic local frame, which then implies the replacement of ordinary differentials with covariant ones.

We begin by discussing the first point in the next subsection and then include the constraints in the following one.

### 3.1    Variational calculus in terms of the first variation functional

As discussed in a previous paper by the author [**1**], it is not necessary to define an action functional in order to obtain the differential equations of an extremal. We briefly summarize the key points of the previous argument.

Suppose that the objects being varied are differentiable functions $x: \mathcal{O} \to M$, where $\mathcal{O}$ is an *r*-dimensional orientable differentiable manifold with boundary and *M* is an *m*-dimensional differentiable manifold. The manifold $\mathcal{O}$ will play the role of a parameter manifold for describing the objects in *M* and we will usually denote local coordinates for it by $t^a$, $a = 1, \ldots, r$; for simplicity, one can think of it as a compact subset of $\mathbf{R}^r$. *M* will represent the configuration manifold of the motion in question.

A *finite variation* of *x* is a differentiable homotopy, i.e., a differentiable map *F*: $[0, 1] \times \mathcal{O} \to M$, $(s, t) \mapsto F(s, t)$ such that $F(0, t) = x(t)$. Although one could define $F(1, t): \mathcal{O} \to M$ to be a finite deformation of the object, since the calculus of variations is only concerned with infinitesimal deformations, one usually does not specify what the final object $F(1, t)$ represents. By partial differentiation of $F(s, t)$, one obtains a vector field on *x*:

$$\delta x(t) = \frac{\partial F(s,t)}{\partial s}\bigg|_{s=0} \tag{3.1}$$

that one calls the *variation* of *x* that is associated with the finite variation *F*. More generally, one regards *any* vector field along *x* as a variation of *x*, although there is clearly the unique finite variation that extends it, in general.

If $(U, x^\mu)$ is a local coordinate system on $U \subset M$ then the components of $\delta x(t)$ with respect to the natural frame field will be denoted by $\delta x^\mu(t)$; that is, they are functions of *t*, not functions of the corresponding point $x(t)$ in *M*. Similarly, if $\mathbf{e}_\mu$ is an anholonomic local frame field on *U* then the components of $\delta x(t)$ will be denoted by $\delta \overline{x}^\mu(t)$.

In the conventional approach to the calculus of variations [**2**, **7**, **8**], one defines a "performance index" on the objects *x* in the form of an *action functional*. This takes the form of the association of a number $S[x]$ to each *x* that one obtains by integrating a Lagrangian density function $\mathcal{L}(t^a, x^\mu, x^\mu{}_{,a})$ over the parameter manifold $\mathcal{O}$:

$$S[x] = \int_{\mathcal{O}} \mathcal{L}(t^a, x^\mu(t), x^\mu_a(t)) dt^1 \cdots dt^r \tag{3.2}$$

The classic variational problem then takes the form of finding the $x$ for which this action function is minimized – or perhaps maximized. Since this is an infinite-dimensional analogue of the minimization for differentiable functions of a finite number of real variables, one proceeds analogously by first looking for the "critical points" of the functional; that is, "points" $x$ of the "infinite-dimensional manifold" of all such objects. One then refers to such critical objects as *extrema* of the action functional. The reason that we are using quotation marks in the preceding remarks is because to actually make those terms rigorous is possible, but not always the most useful approach to the calculus of variations.

When one applies a variation $\delta x$ to $x$ the effect on $S[x]$ defines the *first variation functional*, which, for each $x$, associates the variation $\delta x$ with the number:

$$\delta S|_x[\delta x] = \int_{\mathcal{O}} \left( \frac{\partial \mathcal{L}}{\partial x^\mu} \delta x^\mu + \frac{\partial \mathcal{L}}{\partial x^\mu_a} \frac{\partial (\delta x^\mu)}{\partial t^a} \right) dt^1 \cdots dt^r \,. \tag{3.3}$$

This functional, which is linear in the variations $\delta x$, then plays the role of the "differential" of the functional $S[x]$ at the "point" $x$, and an extremum of $S[x]$ becomes a zero of $\delta S|_x[\delta x]$, at least when one restricts $\delta x$ to some subspace that is defined by the behavior of $\delta x$ on the boundary. However, in order to guarantee that an extremum is actually a minimum, one must move on to the second variation; however, that is beyond the scope of the immediate investigation.

In order to illuminate the most reasonable constraints on the boundary values of $\delta x$, one performs the usual integration by parts to express $\delta S|_x[\delta x]$ as:

$$\delta S|_x[\delta x] = \int_{\mathcal{O}} \frac{\delta \mathcal{L}}{\delta x^\mu} \delta x^\mu \, dt^1 \cdots dt^r + \int_{\partial \mathcal{O}} \left( \frac{\partial \mathcal{L}}{\partial x^\mu_a} \delta x^\mu \right) n_a \mathcal{V}_{\partial \mathcal{O}} \,, \tag{3.4}$$

in which:

$$\frac{\delta \mathcal{L}}{\delta x^\mu} = \frac{\partial \mathcal{L}}{\partial x^\mu} - \frac{\partial}{\partial t^a} \frac{\partial \mathcal{L}}{\partial x^\mu_a} \tag{3.5}$$

is the *variational derivative* of $\mathcal{L}$ with respect to $\delta x$ and $\mathcal{V}_{\partial \mathcal{O}}$ represents the volume element on $\partial \mathcal{O}$, while $n_a$ represents its unit normal [1].

The two basic boundary conditions in $\delta x$ are:

1. *Fixed boundary:*      $\delta x(t) = 0$ then $t \in \mathcal{O}$.

2. *Free boundary:*      $\delta x(t)$ must then satisfy the *transversality condition:*

---

[1] With respect to – say – the metric that is induced from the Euclidian metric on $\mathbf{R}^r$; however, the boundary integrand can be defined in the absence of a metric.

$$\frac{\partial \mathcal{L}}{\partial x^{\mu}_{a}} \delta x^{\mu} = 0 \qquad \text{for all } a = 1, \, \ldots, \, r \tag{3.6}$$

in order for the boundary contribution to the first variation to vanish.

One obtains the differential equations of the extremal in the form:

$$\frac{\delta \mathcal{L}}{\delta x^{\mu}} = 0, \tag{3.7}$$

which are referred to as the *Euler-Lagrange equations.*

Now, one can just as well start the above argument with the first variation functional, suitably generalized. Instead of posing the variational problem as that of minimizing the action functional $S[x]$, one simply poses it the problem of finding the zeroes of the first variation functional, with the same constraints of $\delta x$.

Let us define the first variation function more generally by:

$$\delta S|_{x}[\delta x] = \int_{\mathcal{O}} \left( F_{\mu} \delta x^{\mu} + \Pi^{a}_{\mu} \frac{\partial (\delta x^{\mu})}{\partial t^{a}} \right) dt^{1} \cdots dt^{r}, \tag{3.8}$$

in which $F_{\mu} = F_{\mu}(t^{a}, x^{\nu}, x^{\nu}_{a})$ represent the components of a generalized force density and the $\Pi^{a}_{\mu} = \Pi^{a}_{\mu}(t^{b}, x^{\nu}, x^{\nu}_{b})$ represent the components of a generalized momentum density, which can also represent infinitesimal stresses. By the same integration by parts, one obtains:

$$\delta S|_{x}[\delta x] = \int_{\mathcal{O}} \left( F_{\mu} - \frac{\partial \Pi^{a}_{\mu}}{\partial t^{a}} \right) \delta x^{\mu} dt^{1} \cdots dt^{r} + \int_{\partial \mathcal{O}} \left( \Pi^{a}_{\mu} \delta x^{\mu} \right) n_{a} \mathcal{V}_{\partial \mathcal{O}}, \tag{3.9}$$

If $\delta x(t)$ satisfies the previous boundary conditions that make the second integral vanish then the zeroes of $\delta S|_{x}[\delta x]$ are the $x$ for which:

$$\left( F_{\mu} - \frac{\partial \Pi^{a}_{\mu}}{\partial t^{a}} \right) \delta x^{\mu} = 0 \tag{3.10}$$

for all variations $\delta x(t)$ that satisfy those constraints.

Now, (3.10) is simply a generalization of d'Alembert's principle, which amounts to saying that the *virtual work* $\delta W$ done by the virtual displacement $\delta x$ must vanish in any case, if one regard the divergence in the parenthetical expression as a generalization of the "inertial force" associated with the motion of $x$. Actually, depending upon the nature of $\mathcal{O}$ these equations can just as well be regarded as equations of static equilibrium if one thinks of $F_{\mu}$ as the body forces and $\Pi^{a}_{\mu}$ as the infinitesimal stresses.

The extremal equations that follow from (3.10) are then:

$$F_\mu = \frac{\partial \Pi^a_\mu}{\partial t^a}, \tag{3.11}$$

which then generalize Newton's equations of motion, as well as the equations of elastostatics and elastodynamics.

The path from first variation functionals back to action functionals is defined only in the event that there is a function $\mathcal{L} = \mathcal{L}(t^a, x^\mu, x^\mu_a)$ such that:

$$F_\mu = \frac{\partial \mathcal{L}}{\partial x^\mu}, \qquad \Pi^a_\mu = \frac{\partial \mathcal{L}}{\partial x^\mu_a}. \tag{3.12}$$

As we shall see, this is the requirement a certain 1-form be exact. Hence, the first variation can be defined in mechanical situation in which that 1-form is not exact, such as for forces that are not conservative.

In order to give the foregoing discussion a more modern geometrical context (see, e.g., Saunders [**9**]), we start by observing that the space whose coordinates are $(t^a, x^\mu, x^\mu_a)$ is the manifold $J^1(\mathcal{O}; M)$ of 1-jets of local differentiable functions from $\mathcal{O}$ to $M$, where the 1-jet of $x: \mathcal{O} \to M$ at $t \in \mathcal{O}$ is the equivalence class of all differentiable functions that are defined in some neighborhood of $t$ – and not necessarily the same neighborhood in every case – such that they all take $t$ to $x(t)$ and all have the same differential at $x$ as $dx|_t$ ; one denotes this equivalence class by $j^1_t x$ .

The manifold $J^1(\mathcal{O}; M)$ admits three canonical projections:

| | | |
|---|---|---|
| Source projection: | $J^1(\mathcal{O}; M) \to \mathcal{O},$ | $j^1_t x \mapsto t,$ |
| Target projection: | $J^1(\mathcal{O}; M) \to M,$ | $j^1_t x \mapsto x,$ |
| Contact projection: | $J^1(\mathcal{O}; M) \to \mathcal{O} \times M,$ | $j^1_t x \mapsto (t, x).$ |

The first two of these are not fibrations, but only define a *fibered manifold* structure on $J^1(\mathcal{O}; M)$. That is, although the projections are submersions – so the rank of the differential map is maximal at every point – they do not satisfy the local triviality requirement that is expected of a fiber bundle.

The third projection defines an affine bundle structure over $\mathcal{O} \times M$ the affine space over each $(t, x) \in \mathcal{O} \times M$ takes the form of $T^*_t \mathcal{O} \otimes T_x M$ , since any differential map $dx|_t$ from $T_t \mathcal{O}$ to $T_{x(t)} M$ belongs to that space. The reason that one does not regard it as a vector bundle is that under a change of coordinates on both $\mathcal{O}$ and $M$ the resulting change in the differential matrix $x^\mu_{,a}$ is not linear, but affine, since, under the diffeomorphic replacement of $t^a$ with $\bar{t}^a(t^b)$ and $x^\mu$ with $\bar{x}^\mu(t^a, x^j)$ , which is a diffeomorphism for each value of $t^a$, one must take a total derivative and obtain:

$$\frac{\partial \overline{x}^{\mu}}{\partial \overline{t}^{\,a}} = \left(\frac{\partial \overline{x}^{i}}{\partial t^{b}} + \frac{\partial \overline{x}^{i}}{\partial x^{j}}\frac{\partial x^{j}}{\partial t^{b}}\right)\frac{\partial t^{b}}{\partial \overline{t}^{\,a}}, \tag{3.13}$$

which then generalizes to the transformation rule for the $x_{a}^{i}$ coordinates:

$$\overline{x}_{a}^{i} = \left(\frac{\partial \overline{x}^{i}}{\partial t^{b}} + \frac{\partial \overline{x}^{i}}{\partial x^{j}}\, x_{b}^{j}\right)\frac{\partial t^{b}}{\partial \overline{t}^{\,a}}. \tag{3.14}$$

The reason that we are calling the third projection the "contact" projection is that the elements of the fibers, being linear maps from $T_{t}\mathcal{O}$ to $T_{x}M$, associate tangent subspaces in each $T_{x}M$ with each pair $(t, x)$, in the form of the image of $T_{t}\mathcal{O}$ under the linear map. One refers to such a subspace as a *contact element* and the geometry of jets is sometimes referred to as *contact geometry*, for that reason. In fact, the term "contact geometry" preceded the term "jet manifold" historically[1], and served as the natural geometry for wave mechanics, at least in the geometrical optics approximation.

Of particular interest to us are the *sections* of the source projection, which are differentiable maps $s: \mathcal{O} \rightarrow J^{1}(\mathcal{O}; M)$ that project back to the identity map on $\mathcal{O}$. That is, the 1-jet $s(t)$ always belongs to the fiber over $t$. In local coordinates on $J^{1}(\mathcal{O}; M)$ a section of the source projection will look like:

$$s(t) = (t^{a}, x^{\mu}(t),\ x_{a}^{\mu}(t)). \tag{3.15}$$

Notice that we are not requiring that the coordinates $x_{a}^{\mu}(t)$ be obtained by differentiating the coordinates $x^{\mu}(t)$ with respect to the $t^{a}$. This is because not all sections of the source projection take that form, but only the *integrable* ones. One then says that such an $s$ is the *1-jet prolongation* $j^{1}x$ of a differentiable map $x: \mathcal{O} \rightarrow M$, which one denotes by:

$$s(t) = j^{1}x(t) = (t^{a}, x^{\mu}(t), x_{,a}^{\mu}(t)). \tag{3.16}$$

We can now rephrase the definition of the action functional by making $\mathcal{L}$ a differentiable function $J^{1}(\mathcal{O}; M)$ and pulling it back to a differentiable function on $\mathcal{O}$ by choosing an $x$ and prolonging it:

$$S[x] = \int_{\mathcal{O}} \mathcal{L}(j^{1}x)\mathcal{V}, \tag{3.17}$$

in which $\mathcal{V}$ is the volume element on $\mathcal{O}$.

---

[1] For some classical discussion of contact geometry and wave propagation, one can peruse Vessiot [**10**] or Hölder [**11**], and for a more modern perspective, see Arnol'd [**12**].

In order to define the first variation functional one must prolong the variation $\delta x(t)$, as well as the map $x$. This process is process is similar, in the sense that one adds components that are obtained by differentiation. It is simplest to describe in local coordinates on $J^1(\mathcal{O}; M)$. If $\delta x(t) = \delta x^\mu(t)\partial_\mu$ is a local vector field on $x(t)$ then its 1-jet prolongation is a local vector field on $j^1 x$ that looks like:

$$\delta^1 x(t) = \delta x^\mu(t) \frac{\partial}{\partial x^\mu} + \frac{\partial(\delta x^\mu)}{\partial t^a}(t) \frac{\partial}{\partial x_a^\mu}. \tag{3.18}$$

Although this definition can be globalized, since we shall not use the global expression, we simply refer the curious to Saunders [**9**].

Hence, we can express the first variation functional in the form:

$$\delta\mathcal{S}|_x[\delta x] = \int_{\mathcal{O}} [j^1 x^* \phi(\delta^1 x)]\mathcal{V}, \tag{3.19}$$

if we introduce the *fundamental 1-form* on $J^1(\mathcal{O}; M)$:

$$\phi = F_\mu \, dx^\mu + \Pi_\mu^a dx_a^\mu. \tag{3.20}$$

When this 1-form is exact there will be a smooth function $\mathcal{L}$ on $J^1(\mathcal{O}; M)$ such that $\phi = d\mathcal{L}$. This is, in fact, the case when and only when $\delta\mathcal{S}|_x[\delta x]$ is obtained from an action functional. Hence, the 1-form $\phi$ is the fundamental mechanical definition that replaces the definition of a Lagrangian for the system.

If one regards the points of $J^1(\mathcal{O}; M)$ as kinematical states of the mechanical system in question, while the 1-forms on $J^1(\mathcal{O}; M)$ of the form (3.20) represent dynamical states then the expressions for the components of $f$ as functions of the kinematical states then represent mechanical constitutive laws for the system.

After integrating by parts and taking the boundary conditions on $\delta x$ into account, the first variation functional takes the form:

$$\delta\mathcal{S}|_x[\delta x] = \int_{\mathcal{O}} [j^1 x^* D^* \phi(\delta x)]\mathcal{V}, \tag{3.21}$$

in which we have introduced the 1-form:

$$D^* \phi = \left( F_\mu - \frac{\partial \Pi_\mu^a}{\partial t^a} \right) dx^\mu. \tag{3.22}$$

The operator $D^*$ that takes $\phi$ to $D^* \phi$ then replaces the variational derivative when there is no Lagrangian density. The extremal equations are then expressed simply as:

$$D^*\phi = 0. \tag{3.23}$$

This operator is quite subtle in its manifestation, since it amounts to an adjoint (modulo boundary terms) to the *Spencer operator D* that acts on sections of the source projection. If $s(t)$ is such a section and has the local form that we chose above then:

$$Ds = \left(x_a^\mu - x_{,a}^\mu\right)dt^a \otimes \partial_\mu . \tag{3.24}$$

Hence, $Ds$ vanishes iff $s$ is integrable. This means that the way that one obtains extremal equations from the first variation functional is fundamentally related to the integrability of sections and variations. Although we shall not belabor the details of the integrability issue here, one might confer some of the authors comments on the subject in [**13**] and the references cited therein.

### 3.2    Inclusion of non-holonomic constraints

We already imposed one constraint on the vector space of allowable variations $\delta x(t)$ along a submanifold $x(t)$ in the form of the boundary constraint that is defined by the type of variational problem. Now, let us impose further constraints on allowable variations by requiring that they take their values in some constraint sub-bundle $C(M)$ in $T(M)$; hence, we are imposing linear constraints.

If $C(M)$ is integrable, which corresponds to holonomic linear constraints, then there is a foliation of $M$ by leaves $L$ that have $C(M)$ for their tangent bundle. Since the dimension of $L$ equals the dimension of any fiber of $C(M)$ – i.e., the number of degrees of freedom in the system – any allowable submanifold $x: \mathcal{O} \to M$ must lie within some such leaf. Hence, in the holonomic case it is sufficient to reduce the definition of the variational problem to that of finding extremal submanifolds in the various leaves.

That is, if the constraints are holonomic then there will be a local coordinate system $(U, x^\mu)$ on $U \subset M$ that is adapted to $C(M)$ such that the constraints take the form:

$$\delta x^\alpha = 0, \qquad \alpha = N+1, \ldots, m. \tag{3.25}$$

The effect of this on the extremal equations is to reduce the integrand of the first variation functional to:

$$D^*f_i \, \delta x^i = (F_i - \partial_a \Pi_i^a)\delta x^i. \tag{3.26}$$

The constrained extremal equations then become simply:

$$F_i - \partial_a \Pi_i^a = 0. \tag{3.27}$$

Furthermore, the constraint implies that the extremal solution $x$ will involve only the coordinates $x^i$, $i = 1, \ldots, N$ that lie within a particular leaf of the constraint foliation. Hence, there has been a reduction of dimension in the extremal problem from $m$ to $N$.

If $C(M)$ is not integrable, which corresponds to non-holonomic linear constraints, then this situation is no longer true. However, one can simplify the situation by representing the fibers of $C(M)$ locally using an adapted anholonomic local frame field $\mathbf{e}_\mu$ on $U \subset M$ that makes the vectors $\mathbf{X}$ of $C(U)$ look like $\mathbf{X} = X^i \mathbf{e}_i$ or $\theta^\alpha(\mathbf{X}) = X^\alpha = 0$, $\alpha = N+1, \ldots, m$.

In order to include the constraints into the variational problem, all that we really need to do is to express the components of our various geometric objects in terms of this anholonomic local frame field and its reciprocal coframe field, whereas so far we have only used holonomic frame fields.

First, we must address the change in the derivative coordinates under the change of local frame field on $M$. If $\mathbf{e}_\mu = A_\mu^\nu \partial_\nu$ and $\theta^\mu = \tilde{A}_\nu^\mu dx^\nu$ then one has:

$$x_a^\mu (dt^a \otimes \partial_\mu) = \overline{x}_a^\mu (dt^a \otimes \mathbf{e}_\mu), \qquad (3.28)$$

which makes:

$$\overline{x}_a^\mu = \tilde{A}_\nu^\mu x_a^\nu. \qquad (3.29)$$

This is not to be confused with the *coordinate* transformation of $x_a^\mu$ that follows from replacing $t^a$ with $\overline{t}^b$ and $x^i$ with $\overline{x}^i(t, x)$, since we are not actually altering the basic coordinates.

By differentiation, one then finds:

$$dx_a^i = A_\nu^\mu \nabla \overline{x}_a^\nu. \qquad (3.30)$$

Dually, one finds for the transformation of the local vector fields $\partial / \partial x_a^i$ that in order to preserve the reciprocal relationship between them and $dx_a^i$ one must replace them with:

$$\mathbf{e}_\mu^a = A_\mu^\nu \frac{\partial}{\partial x_a^\nu}, \qquad (3.31)$$

which then makes:

$$\nabla \overline{x}_a^\mu(\mathbf{e}_\nu^b) = \delta_\nu^\mu \delta_b^a. \qquad (3.32)$$

When one combines this with $dx^i = A_\nu^\mu \theta^\nu$, one finds that the fundamental 1-form $\phi$ becomes:

$$\phi = \overline{F}_\mu \theta^\mu + \overline{\Pi}_\mu^a \nabla \overline{x}_a^\mu = (\tilde{A}_\mu^\nu \overline{F}_\nu dx^\mu + \tilde{A}_\nu^\mu \overline{\Pi}_\mu^a dx_a^\mu) = F_\mu dx^\mu + \Pi_\mu^a dx_a^\mu. \qquad (3.33)$$

Thus:

$$\overline{F}_\mu = A_\mu^\nu F_\nu, \qquad \overline{\Pi}_\mu^a = A_\mu^\nu \Pi_\nu^a. \qquad (3.34)$$

One must, of course, notice that the effect of introducing covariant differentials in place of ordinary ones is to allow one to use the simpler rules for the transformation of components from holonomic to anholonomic frames by absorbing the inhomogeneous part of the transformation into the covariant differential.

Dually, one finds that the 1-jet prolongation of $\delta^1 x$ becomes:

$$\delta^1 x = \delta \overline{x}^\mu \mathbf{e}_\mu + (\nabla_a \delta \overline{x}^\mu) \mathbf{e}_\mu^a \,, \tag{3.35}$$

which gives:

$$\delta \overline{x}^\mu = \tilde{A}_\nu^\mu \delta x^\nu \,, \qquad \nabla_a \delta \overline{x}^\mu = \tilde{A}_\nu^\mu \partial_a \delta x^\nu \,. \tag{3.36}$$

Hence, the virtual work integrand in the first variation functional becomes:

$$\phi(\delta^1 x) = \overline{F}_\mu \delta \overline{x}^\mu + \overline{\Pi}_\mu^a \nabla_a \delta \overline{x}^\mu \,, \tag{3.37}$$

which, when one takes into account the product rule for the covariant derivative $\nabla_a$, becomes:

$$\phi(\delta^1 x) = D^* \phi(\delta \overline{x}) + \nabla_a (\overline{\Pi}_\mu^a \delta \overline{x}^\mu) \,, \tag{3.38}$$

in which we now have:

$$D^* \phi = \left( \overline{F}_\mu - \nabla_a \overline{\Pi}_\mu^a \right) \theta^\mu \,. \tag{3.39}$$

Since the second term on the right-hand side of (3.38) becomes a boundary term under integration, the extremal equations now take the form of:

$$\left( \overline{F}_\mu - \nabla_a \overline{\Pi}_\mu^a \right) \delta \overline{x}^\mu = 0 \tag{3.40}$$

for every variation $\delta^1 x$ that is consistent with the constraints that were imposed by $C(M)$. Of course, the advantage of using anholonomic local frame fields and covariant differentiation is that all one needs to do is notice that in such a local frame field the constraint on $\delta x$ is simply that $\delta \overline{x}^\alpha = 0$ for $\alpha = N+1, \ldots, m$. Since this leaves the corresponding parenthetical expression (3.40) arbitrary, the ultimate effect of imposing non-holonomic constraints is to give the extremal equations in the form:

$$0 = D^* \phi_i = \overline{F}_i - \nabla_a \overline{\Pi}_i^a \,, \qquad i = 1, \ldots, N. \tag{3.41}$$

The difference between this result and the reduction of dimension that would follow from a holonomic constraint is in the fact that since $\theta^i$, $i = 1, \ldots, N$ do not represent coordinate differentials, the integration of (3.41) to obtain a constrained extremal $x$ must still involve an $x$ whose local coordinates in $M$ involve all $m$ of the unconstrained coordinates $x^\mu$, and not just the first $N$ of them, namely, the $x^i$.

### 4       Example: disc rolling on a plane

A simple example of a mechanical system that is subject to ideal non-holonomic constraints is given by a vertical disc, such as an ideal tire, rolling without slipping on a plane in the absence of dissipative forces. We illustrate this situation in Fig. 1 in order to define the configuration manifold $M$ of the system.

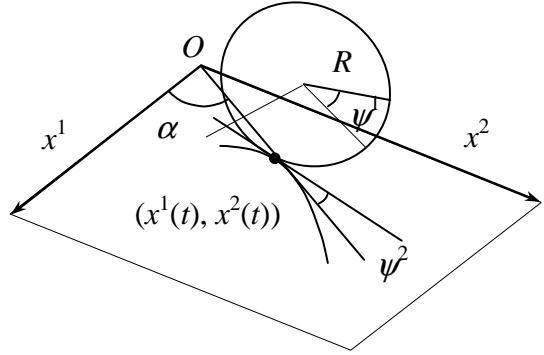

Figure 1.  An ideal vertical disc rolling without slipping on a plane.

In this situation, our parameter manifold will be simply $\mathcal{O} = [0, 1]$, since we will be concerned with 1-jets of curve segments; i.e., point mechanics. The configuration manifold is four-dimensional and takes the form of $M = T^2 \times \mathbf{R}^2$, whose local coordinates are given by $(\psi^1, \psi^2, x^1, x^2)$. A differentiable curve in $M$ then takes the form of

$$x(t) = (\psi^1(t), \psi^2(t), x^1(t), x^2(t)). \tag{4.1}$$

Hence, in the natural frame field that goes with such a coordinate system the velocity vector field along $x(t)$ takes the form:

$$\mathbf{v}(t) = \omega^i \frac{\partial}{\partial \psi^i} + v^\alpha \frac{\partial}{\partial x^\alpha}, \qquad (\omega^i = \frac{d\psi^i}{dt}, \qquad v^\alpha = \frac{dx^\alpha}{dt}). \tag{4.2}$$

The jet manifold $J^1(\mathcal{O}; M)$ then has local coordinates of the form $(t, \psi^j, x^\alpha, \omega^j, v^\alpha)$.

In order to describe the non-holonomic constraint, it is simplest to first transform to polar coordinates in the $x^1 x^2$-plane. With $r$ defined to be $[(x^1)^2 + (x^2)^2]^{1/2}$, one has:

$$x^1 = r \cos \alpha, \quad x^2 = r \sin \alpha. \tag{4.3}$$

Differentiation of these equations, under the assumption that all variables are functions of $t$, gives:

$$\dot{x}^1 = \dot{r}\cos\alpha - r\dot{\alpha}\sin\alpha, \qquad \dot{x}^2 = \dot{r}\sin\alpha + r\dot{\alpha}\cos\alpha. \qquad (4.4)$$

One then derives the moving coframe $\{dr, rd\alpha\}$ for the description of velocity vectors from:

$$\begin{bmatrix} \dot{x}^1 \\ \dot{x}^2 \end{bmatrix} = R(-\alpha)\begin{bmatrix} \dot{r} \\ r\dot{\alpha} \end{bmatrix}, \qquad R(-\alpha) = \begin{bmatrix} \cos\alpha & -\sin\alpha \\ \sin\alpha & \cos\alpha \end{bmatrix}. \qquad (4.5)$$

However, the moving frame $\{\mathbf{\varepsilon}_1, \mathbf{\varepsilon}_2\}$ that we ultimately want is obtained by a further rotation in the $x^1 x^2$-plane through $\psi^2$ that brings the $\mathbf{\varepsilon}_1$ vector parallel to the plane of the disc while the $\mathbf{\varepsilon}_2$ vector is perpendicular to it in the right-hand sense. Combining the two rotations gives:

$$\mathbf{\varepsilon}_\alpha = R_\alpha^\beta(-\alpha - \psi^2)\partial_\beta, \qquad \xi^\alpha = R_\beta^\alpha(\alpha + \psi^2)dx^\beta. \qquad (4.6)$$

The constraint of rolling without slipping then takes the simple form:

$$\xi^1(\mathbf{v}) = v^\parallel = R\omega^1, \qquad \xi^2(\mathbf{v}) = v^\perp = 0. \qquad (4.7)$$

Hence, we can define an adapted non-holonomic local coframe field $\{d\psi^j, \theta^\alpha\}$, in which:

$$\theta^\alpha \equiv \xi^\alpha - a_j^\alpha d\psi^j, \qquad a_i^\alpha = \begin{bmatrix} R & 0 \\ 0 & 0 \end{bmatrix} \qquad (4.8)$$

defines the non-holonomic constraint by way of

$$\theta^\alpha(\mathbf{v}) = v^\alpha = 0. \qquad (4.9)$$

We then combine all three frame transitions into:

$$\begin{bmatrix} d\psi^1 \\ d\psi^2 \\ \theta^1 \\ \theta^2 \end{bmatrix} = A_\nu^\mu \begin{bmatrix} d\psi^1 \\ d\psi^2 \\ dx^1 \\ dx^2 \end{bmatrix}, \qquad (4.10)$$

with:

$$A_\nu^\mu = \begin{bmatrix} \delta_j^i & 0 \\ \hline -a_j^\alpha & R_\beta^\alpha(\alpha + \psi^2) \end{bmatrix}, \qquad \tilde{A}_\nu^\mu = \begin{bmatrix} \delta_j^i & 0 \\ \hline R_\beta^\alpha(-\alpha - \psi^2)a_i^\beta & R_\beta^\alpha(-\alpha - \psi^2) \end{bmatrix}. \qquad (4.10)$$

Note that this means our non-holonomic frame field is not actually semi-holonomic, in this case.

We then find that:

$$dA^\mu_\nu = \begin{bmatrix} 0 & \vdots & 0 \\ \hline 0 & \vdots & dR^\alpha_\beta(\alpha + \psi^2) \end{bmatrix},$$
(4.11)

so:

$$\gamma^\mu_\nu = \begin{bmatrix} 0 & \vdots & 0 \\ \hline dR^{\tilde\alpha}_\beta a^\beta_j & \vdots & -dR^\alpha_\gamma \tilde R^\gamma_\beta \end{bmatrix}.$$
(4.12)

We then obtain the specific components of the desired connection in the form:

$$\gamma^i_j = \gamma^i_\alpha = 0, \qquad \gamma^\alpha_j = \begin{bmatrix} 0 & 0 \\ R & 0 \end{bmatrix}(d\alpha + d\psi^2), \qquad \gamma^\alpha_\beta = \begin{bmatrix} 0 & -1 \\ 1 & 0 \end{bmatrix}(d\alpha + d\psi^2).$$
(4.13)

We then have to express the differential $d\alpha$ in terms of the coframe $\{d\psi^1, d\psi^2, \theta^1, \theta^2\}$. By tedious computation, we obtain:

$$d\alpha = \frac{1}{r}\tilde R^2_\beta(\psi^2)(a^\beta_i d\psi^i + \theta^\beta) = \frac{R}{r}\sin\psi^2 d\psi^1 + \frac{1}{r}\sin\psi^2\theta^1 + \frac{1}{r}\cos\psi^2\theta^2,$$
(4.14)

which can then be substituted into (4.13).

The way that we obtain the constrained equations of motion then begins with the specification of the fundamental 1-form $\phi$ in the holonomic coframe:

$$\phi = \tau_i\, d\psi^i + F_\alpha\, dx^\alpha + L_i\, d\omega^j + p_\alpha\, dv^\alpha.$$
(4.15)

Under the transition to the non-holonomic coframe $\{d\psi^j, \theta^\alpha\}$, this will take the form:

$$\phi = \bar\tau_i d\psi^i + \bar F_\alpha\theta^\alpha + \bar L_i d\omega^j + \bar p_\alpha\nabla\bar v^\alpha,$$
(4.16)

in which the $\bar v^\alpha$, $\alpha = 1, 2$ are the last two components of the velocity vector in the $\mathbf{e}_\mu$ frame. Hence:

$$\nabla\bar v^\alpha = A^\alpha_\mu dv^\mu = A^\alpha_i d\omega^j + A^\alpha_\beta dv^\beta.$$
(4.17)

We then compute $dv^\alpha$ from the inverse transformation:

$$dv^\alpha = \tilde A^\alpha_\mu\nabla v^\mu = \tilde A^\alpha_j d\omega^j + \tilde A^\alpha_\beta\nabla\bar v^\beta.$$
(4.18)

Substituting this into (4.15), along with the expression for $\theta^\alpha$ that one obtains from (4.10), along with some calculation, gives the transformation of the components:

$$\overline{\tau}_i = \tau_i + \tilde{A}_i^\alpha F_\alpha, \qquad \overline{F}_\alpha = \tilde{A}_\alpha^\beta F_\beta, \qquad \overline{L}_i = L_i + \tilde{A}_i^a p_\alpha, \qquad \overline{p}_\alpha = \tilde{A}_\alpha^\beta p_\beta. \quad (4.19)$$

One immediately recognizes that these equations become much easier to work with if one uses the second and fourth of them in the first and second of them, respectively:

$$\overline{\tau}_i = (\tau_1 + \overline{F}_1 R, \tau_2), \qquad \overline{L}_i = (L_1 + \overline{p}_1 R, L_2). \quad (4.19)$$

These expressions now an immediate physical meaning: The total torque about the transverse axis of the disc consists of the applied torque plus the moment of the tangential applied force that acts on the center of mass, while the total angular momentum about that axis consists of both the "intrinsic" part:

$$L_1 = I_1 \omega_1, \quad (4.20)$$

in which $I_1$ is the moment of inertia about that axis, plus an "orbital" part that comes from the moment of the momentum of the center of mass:

$$\overline{p}_1 R = m \overline{v}_1 = m R \ \omega_1, \quad (4.21)$$

in which $m$ is the mass of the disc; we have implicitly lowered the velocity indices with the Euclidian metric.

The torque about the vertical axis of the disc, as well as its angular momentum, remain unaffected by the transformation:

$$\overline{\tau}_2 = \tau_2, \qquad \overline{L}_2 = L_2 = I_2 \omega_2, \quad (4.22)$$

in which $I_2$ is the moment of inertia about the vertical axis.

The general form of the equations of motion in this anholonomic coframe is:

$$\overline{F}_\mu = \nabla_t \overline{p}_\mu = \frac{d\overline{p}_\mu}{dt} - c_\mu^\nu(\mathbf{v}) \overline{p}_\nu, \quad (4.23)$$

however, the constraints imply that there no motion in the transverse direction, so these equations reduce to:

$$\overline{\tau}_i = \frac{d\overline{L}_i}{dt} - c_i^j(\mathbf{v}) L_j - c_i^\alpha(\mathbf{v}) \overline{p}_\alpha = \frac{d\overline{L}_i}{dt}, \quad \overline{v}^\alpha = 0. \quad (4.24)$$

With the aforementioned substitutions, the first set of equations becomes:

$$\frac{d\omega_1}{dt} = \frac{\tau_1(t) + R\overline{F}_1(t)}{I_1 + mR^2}, \qquad \frac{d\omega_2}{dt} = \frac{1}{I_2} \tau_2(t). \quad (4.25)$$

As long as the applied force and torque as known as functions of time, the initial-value problem for these equations can be integrated by quadratures, in principle. Similarly, one can integrate the expressions for $\omega_i(t)$ to obtain the angles $\psi_i(t)$.

One can then use the expressions for $\omega_i(t)$ in the second set of equations in (4.24) to obtain $dx^i/dt$ by quadratures, as well, by taking into account the rolling-without-slipping constraint in order to go from angular velocity to linear velocity.

## 5    Discussion

So far, we have seen that by starting with the first-variation functional one can make a natural and straightforward variational formulation of mechanical systems that are either non-conservative or subject to non-holonomic constraints. The picture that emerges is that starting with an action functional is only advisable when the fundamental 1-form that we defined above is an exact form, which implies that both the generalized force term and the generalized momentum term must be exact.

Although one usually assumes that the generalized momentum is associated with a generalized kinetic energy in some way, this tends to be a consequence of starting with notions that are rooted in point mechanics, rather than continuum mechanics. Hence, the possibility that a momentum 1-form might have – say – a non-vanishing exterior derivative as a function of velocity seems remote, since that suggests that the generalized mass density would have to a function of velocity, as well; of course, that makes perfect sense in relativistic dynamics, of one considers relative mass densities. Thus, it might be worthwhile to examine the role of inexactness in the generalized momentum 1-form.

Another aspect of the present formalism that seems to promise considerable applications is in the formulation of open mechanical systems. In such systems, there is some well-defined decomposition of the system into a "internal" and "external" sub-system, although the external sub-system might very well take the form of unmodeled phenomena at the atomic-to-subatomic level. Instead of conservation laws and equilibrium, one must consider balance laws and perturbations about equilibrium, which naturally suggests the issue of the stability of extrema.

In order to account for the issue of stability of extrema, one must re-examine the second variation. Clearly, one cannot compute the usual Hessian in the absence of a Lagrangian, but one can use more general methods for examining the local behavior in the neighborhood of zeroes of 1-forms.

The subject of open systems seems to have considerable application to the variational methods of quantum physics. For instance, a recurring theme in most discussions of the Feynman path integral is the fact that non-extremal paths contribute to the total transition probability from one state to another. Furthermore, one always thinks of the mechanics of quantum systems as being non-trivially affected by external influences in the form of either measurements performed on the system or the unmodeled internal states, such as the creation and annhilation of virtual matter-antimatter pairs during the propagation of waves, whether in the form of photons or matter waves.